\documentclass[aip,
 amsmath,amssymb,
reprint,%
]{revtex4-1}

\usepackage{graphicx}
\usepackage{dcolumn}
\usepackage{bm}

\usepackage[utf8,latin1]{inputenc}
\usepackage[T1]{fontenc}
\usepackage{mathptmx}
\usepackage{etoolbox}

\makeatletter
\def\@email#1#2{%
 \endgroup
 \patchcmd{\titleblock@produce}
  {\frontmatter@RRAPformat}
  {\frontmatter@RRAPformat{\produce@RRAP{*#1\href{mailto:#2}{#2}}}\frontmatter@RRAPformat}
  {}{}
}%
\makeatother
\begin{document}

\title[Minimally driven Kapitza oscillator with Newtonian mechanics and geometry]{Minimally driven Kapitza oscillator: A pedagogical perspective from Newtonian mechanics and geometry}
\author{Mainak Pal}
\affiliation{Department of Physics, University of Florida}

\begin{abstract}
The unstable top-equilibrium point of a simple pendulum turns stable when its pivot point is given a fast and strong enough vertical vibration.
Known as the Kapitza oscillator, it has {four symmetrically spaced points of equilibrium} in {absence of gravity}, out of which {two are stable} and {two are unstable}.
This article, completely based on a {geometric argument} and an {elementary intuition in Newtonian mechanics}, is a visual and pedagogical
exposition of (a) {why} the oscillator has four symmetrically spaced equilibrium points in absence of gravity, (b) {which} of them are stable
or unstable, (c) {why} they are so and (d) {how} the stability and position and number of the equilibrium points
change when gravity is turned on gradually along the line of vibration of the pivot of the oscillator. A minimal impulsive drive of the pivot is sufficient to illustrate the bare bones of the phenomenon. I propose a construction that can sustain the minimal drive passively in absence of dissipative forces, or actively if all dissipative forces can't be eliminated. In either of the cases, the discussed arguments apply.
\end{abstract}

\maketitle
\section{Introduction}
A simple pendulum is a most elementary physical system that embodies the simple harmonic motion, a dynamics of fundamental importance and often a method of approximation in multiple branches of physics, so much so that Sidney Coleman once said ``The career of a young theoretical physicist consists of treating the harmonic oscillator in ever-increasing levels of abstraction''. When a simple pendulum is augmented with some other elementary constructs, this deceptively simple system oftentimes shows extremely rich and fascinating physical behaviors. 
When pivoted onto the end of another simple pendulum, to make a double pendulum, it demonstrates chaotic behavior \cite{doublependulum}, which is a wonderful and important property of many dynamical systems, and is abundant in nature and everyday life around us. 

Another scenario, which we'll discuss pedagogical insights about, arises when the pivot point of a simple pendulum is vibrated vertically. Known as the Kapitza oscillator, so since Pyotr Kapitza first provided an explanation of its dynamics with Lagrangian mechanics,
this vibrating-pivot oscillator is perplexing at first sight, because the unstable top-equilibrium point of a simple pendulum becomes a stable equilibrium point as a result of the periodic drive of the pivot. The stable bottom-equilibrium point continues to be stable, in fact becomes stabler, showing a higher frequency of oscillation, as compared to the `simple non-vibrating-pivot pendulum'. This system, like the double pendulum, also shows a variety of rich behaviors and is discussed actively not only in classical non-linear systems, but also in driven condensed matter systems, in its quantum version \cite{Richards2018,PhysRevLett.126.253601,PhysRevB.98.224305}.

Treatment of the Kapitza osillator in literature \cite{Landau1976Mechanics}, including Kapitza's original work \cite{Kapitza1} is exclusively based on Lagrangian framework of classical mechanics. An analytic description of the phenomenon based on Newtonian mechanics is surprisingly scarce \cite{doi:10.1119/1.1365403} and needs to be augmented with a lucid, pictorial and pedagogical treatment. Moreover, the vibration of the pivot in 
 the literature is taken sinusoidal, i.e. continuously varying in time rather than taking discrete values. This leaves the explanation based on Newtonian mechanics proposed in earlier literature in a bit of ambiguity, and puts forward the question, ``What could be the simplest/minimal possible periodic drive of the pivot that can stabilize the upward equilibrium position and still admits a simple visual explanation?''. In this article I consider, for pedagogical purpose, the vibration created by a sequence of oppositely directed instantaneous impulsive forces. Approximating quantities that are continuous in time, as a sequence of delta functions, often  significantly helps to understand the crux of the physical phenomena (see, for example, Feynman's treatment of gravitational force \cite{goodstein1996feynman} as a series of delta functions or impulsive forces, while considering elliptical path of one mass in the gravitational field of another). I discuss that such a minimal and simple form of driving force can also provide stability to the top-equilibrium point, and moreover, admits a clear real-time intuition into the mechanism of the phenomenon based on a simple principle of Newtonian mechanics and elementary geometry. At first, I restrict the discussion to only a driven pendulum without gravity for clarity of understanding, and later explain, {with the same principle}, the qualitative effect of gravity on the equilibrium positions.

\section{Pendulum on a Table: No pivot-vibration and no gravity}
\begin{figure}[tb]
    \centering
    \vspace{0.5cm}
    \includegraphics[scale=1.2]{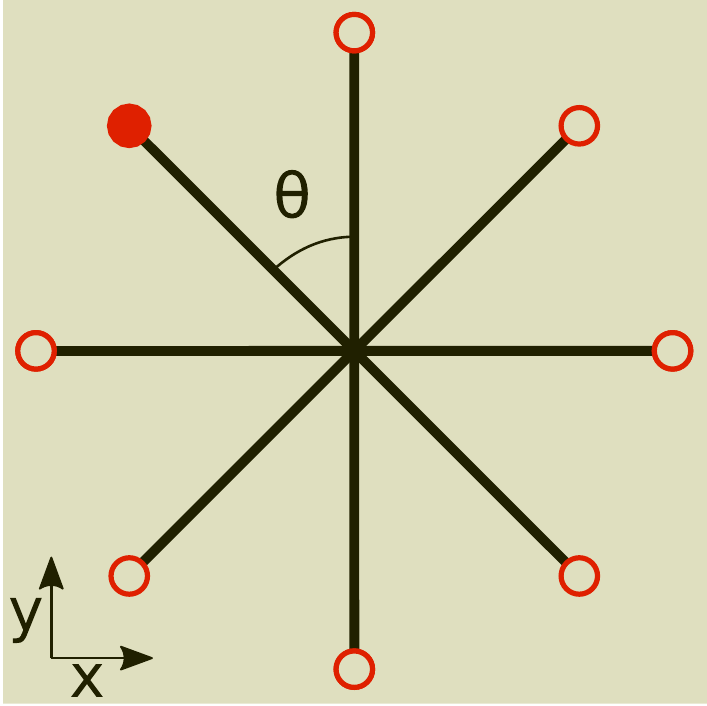}
    \caption{Various angular positions of a simple pendulm/oscillator on a horizontal friction-less table. The black line with filled red circle shows an actual position of the pendulum and the ones with open circles show some of its possible positions. All angular positions 
    are in neutral equilibrium.}
    \label{fig:fig1}
\end{figure}

Imagine a simple pendulum without any vibration of the pivot in absence of gravity. One way of taking the effect of gravity off is restricting the motion of the pendulum on a horizontal friction-less plane.
This is the most elementary scenario imaginable for the pendulum (see Fig. \ref{fig:fig1}).
As is obvious, all angles are positions of neutral equilibrium in this state of the pendulum.

\section{Switch on pivot-vibration}
However, as soon as a driving force starts vibrating the pivot, the dynamics becomes non-trivial, and most of the positions of equilibrium are eliminated. 
\begin{figure*}
    \centering
    \vspace{0.5cm}
    \includegraphics[scale=0.15]{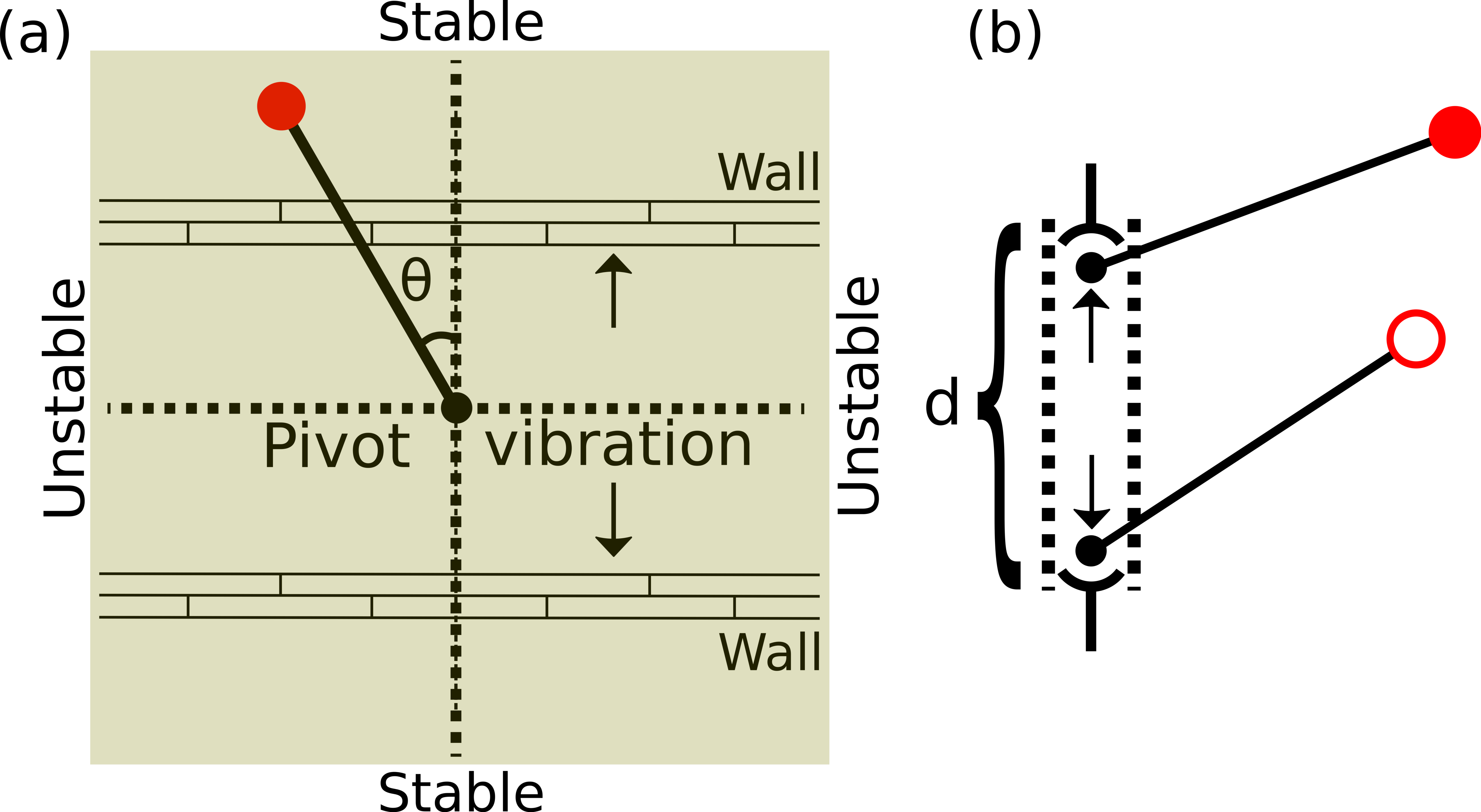}
    \caption{The vibrating-pivot oscillator in absence of gravity. It has four symmetrically distributed positions of equilibrium. Two of them are stable and two are unstable. (a) The pivot (black dot) is confined between two walls. The plane that contains the pivot and the wall, is slightly below but parallel to, the plane of movement of the pendulum. (b) More compact and activated version of (a). The pistons at the end of a vertical channel, that confines the pivot, can either stay static like walls to reflect the pivot, or can themselves vibrate with tiny amplitude to provide impulsive drive to the pivot. The black line with filled red dot shows the real pendulum and the one with open red circle shows a possible position of the pendulum. $d$ (shown not to scale) is large compared to the pistons' amplitude of vibration, but small compared to the pendulum's amplitude of oscillation. }
    \label{fig:fig2}
\end{figure*}
This scenario is presented in Fig. \ref{fig:fig2}. The pivot is denoted by a small black dot. One can of course apply periodic impulsive forces up and down in sequence, with some active external agent. However, this can be arranged relatively passively as well, by confining the pivot between two slightly separated immovable walls and firing the pivot normally to one of these walls (see Fig.\ \ref{fig:fig2} a). All surfaces are assumed frictionless, and all collisions are assumed completely elastic. The pivot should keep on getting reflected between the walls, which impart the oppositely directed sequence of impulsive forces to the pivot. The `plane that contains the walls and the pivot' can be different than but parallel to the `plane of motion of the oscillator', such that the rotation of the oscillator is not hindered by the walls. A more compact apparatus for pivot vibration is presented in Fig.\ \ref{fig:fig2} b, which shows two pistons at two ends of a channel that confines the pivot. The two pistons can be kept immovable like the walls, and the pivot can be fired towards any one of the pistons. Alternatively, the pistons can be vibrated themselves to produce the impulsive forces and sustain the pivot-vibration, in case all dissipative forces can't be eliminated. 
Vibration amplitude of the pistons is much smaller than the vibration amplitude $d$ of the pivot, which in turn is much smaller than the oscillation amplitude of the pendulum. For clarity of presentation, $d$ has been shown disproportionately large in the diagrams.

As a result of the pivot-vibration, the equilibrium positions that are parallel to the direction of vibration, transition to stable from neutral, whereas the equilibrium positions perpendicular to the direction of vibration transition from neutral to unstable. In the following sections we'll see why and how.    
\subsection{A stick, inclined more, rotates faster when struck}
\label{RotatesMore}
The crux of the explanation rests upon one simple observation. When a simple stick is struck with a force, whether the stick will develop any rotational motion, depends on whether the direction of force passes through its centre of mass. Hence, if a stick is struck at one of its ends in a direction parallel to its length, the stick doesn't gain any angular velocity. However, if the stick has some inclination angle with the direction of the force, it {will} develop an angular velocity. The {\it developed angular velocity is more, if the angle of inclination is more, irrespective of the sense of rotation}. This is simply because of the fact that although the acting force is {same}, the {arm of the torque is more if the inclination is more}. As we'll see, this simple piece of information/observation is sufficient to show that as soon as vibration of the pivot starts, infinitely many directions of neutral equilibrium of the stick/pendulum disappear, leaving only four, out of which two are stable, and two are unstable.
\subsection{Explaining the stable equilibrium points}
\begin{figure*}
    \centering
    \includegraphics[scale=0.037]{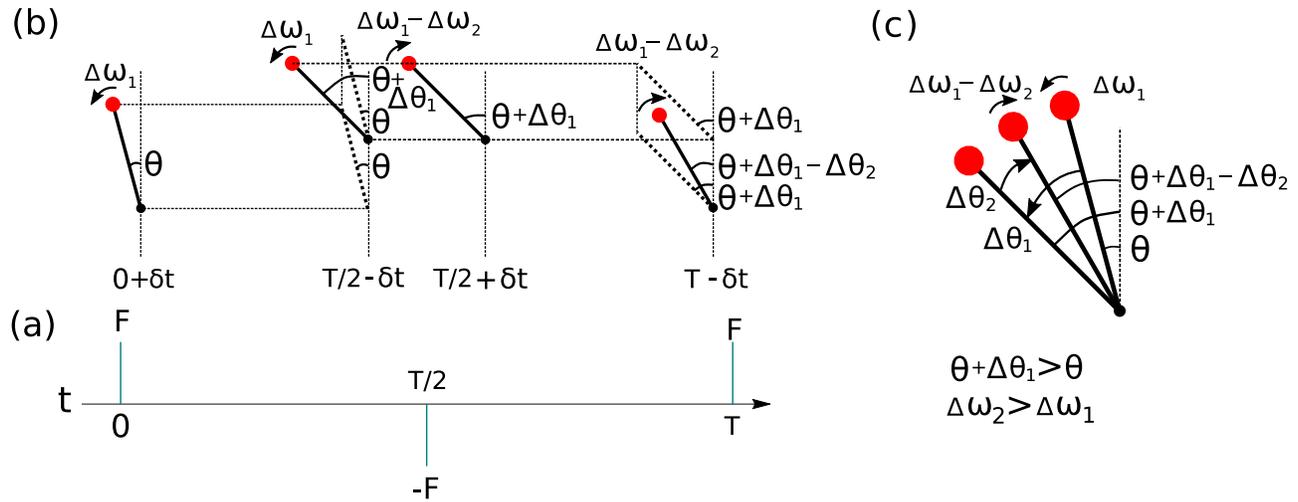}
    \caption{During any full-cycle $T$ of pivot-vibration, the oscillator gains an incremental angular velocity towards the axis of vibration, clockwise if $\theta>0$, anticlockwise if $\theta<0$. (a) shows the impulsive forces by the pistons/walls on the pivot in intervals of $T/2$, (b) shows the incremental angular velocities at the start and the end of these half-cycles of duration $T/2$, (c) shows these positions of the pendulum at one place to facilitate comparison.} 
    \label{fig:fig3}
\end{figure*}

{In the beginning of the motion, the oscillator may have an anticlockwise zero or non-zero velocity $\omega$. If non-zero, this $\omega$ would try to drive the oscillator away from the top-equilibrium position. The angular velocities and displacements that I'll discuss below, and show in Fig.\ \ref{fig:fig3}, do not include this background angular velocity $\omega$, and the displacement accumulated due to that $\omega$. In fact, I discuss only the details of the dynamics due to pivot vibration, and show how that dynamics decreases the background angular velocity $\omega$, eventually turning it negative. Hence, the oscillator will eventually come back to the top-equilibrium point.} 

Let's consider the pivot as our frame of reference. The angular velocities, that we'll consider in this article, are around the pivot, and taken positive if anticlockwise. The angular positions are measured from the upward axis of pivot-vibration, and taken positive, if anticlockwise. The oscillator is at an angular position $\theta_{}$ at $t=0-\delta t$, just before the beginning of the pivot-vibration.
Here $\delta t$ is an infinitesimal time interval. 
An upward impulsive force $F$ acts on the pivot at $t=0$, when the pivot hits the lower piston.
{\it Because} $\theta_{}$ is non-zero, the oscillator gains an anticlockwise angular velocity $\Delta \omega_{1}$ at $t=0+\delta t$ (see Fig.\ \ref{fig:fig3}).
At $t=T/2$, the oscillator
is acted upon by the opposite impulsive force $-F$ from the upper piston.
As a result of this impulse, the oscillator gains at $t=T/2+\delta t$, a clockwise angular velocity $-\Delta\omega_{2}$. {\it Because} $\theta+\Delta\theta_{2} > \theta_{}$, we have $\Delta\omega_{2}>\Delta\omega_{1}$ by the principle described in \ref{RotatesMore}. 
So, at the end of one time-period of pivot-vibration $t=T-\delta t$, but just before hitting the lower piston, the oscillator has gained an angular velocity $\Delta\omega_1-\Delta\omega_2 < 0$. {\it Hence, the background anticlockwise angular velocity of the oscillator, $\omega$ decreases by $\Delta\omega_1-\Delta\omega_2$, making the oscillator decelerate.}
Now, the same thing continues for subsequent time-periods. 
The two half-cycles of pivot-vibration within each time-period provide an incremental { clockwise} angular velocity, $\Delta\omega_3-\Delta\omega_4 < 0$, $\Delta\omega_5-\Delta\omega_6 < 0$, ..., $\Delta\omega_{2n-1}-\Delta\omega_{2n} < 0$ and so on. The pivot-vibration needs to be sufficiently faster than the pendulum oscillation so that the accumulated decrement in angular velocity $(\Delta\omega_1-\Delta\omega_2) + (\Delta\omega_3-\Delta\omega_4) + ... + (\Delta\omega_{2n-1}-\Delta\omega_{2n}) < 0$ after $n_0$ pivot-vibrations makes $\omega$ negative, and the pendulum starts coming back towards the top-equilibrium point. Note that the net angular displacement in one  full-cycle, $\Delta \theta_1 - \Delta \theta_2$ immediately after the pivot-vibration starts, is not necessarily negative. This would take until a few cycles of vibration $n_0$, when 
\begin{eqnarray*}
 \bigg|\sum_{n=1}^{n_0+1} (\omega_{2n-1} - \omega_{2n})\bigg| > \bigg|\sum_{n=1}^{n_0} (\omega_{2n-1} - \omega_{2n}) + \omega_{2n+1}\bigg|   
\end{eqnarray*}
is satisfied. 

It's not difficult to see that if the impulsive force is, downward in the first half-cycle of pivot vibration, and upward in the second half-cycle, then also the same logic as in \ref{RotatesMore} applies, and there would be a net decrement in the background angular velocity $\omega$ after a full-cycle of pivot-vibration.
   
\subsection{Explaining unstable equilibrium points}
Now, if the background angular velocity $\omega$ of the oscillator is zero and initial angular position of the oscillator at $t=0$ is $ \theta=\pi/2-\Delta\theta$, such that at $t=T/2$ the angular position is exactly $\pi/2+\Delta\theta$, then $\Delta\omega_{1}=\Delta\omega_{2}$ using the logic in \ref{RotatesMore}. That is, there is no net gain or loss in angular velocity after any full-cycle of pivot vibration. Hence, the oscillator keeps on taking two angular positions $\pi/2\pm\Delta\theta$ alternately, which means that the angular positions $\theta=\pm\pi/2$ are equilibrium positions as well. However, if the initial angular position is slightly different, say $\theta=\pi/2-\Delta\theta^{\prime}$, such that at $t=T/2$ the angular position is $\pi/2+\Delta\theta^{\prime\prime}$ with $\Delta\theta^{\prime}\neq\Delta\theta^{\prime\prime}$, then $\Delta\omega_{1}\neq\Delta\omega_{2}$, and the oscillator will gradually start moving towards either  $\theta=0\ {\rm or}\ \pi$. Hence, certainly $\theta=\pm\pi/2$ are unstable points of equilibrium.
\section{Effect of Gravity}
\begin{figure*}
    \centering
    \vspace{0.5cm}
    \includegraphics[scale=0.15]{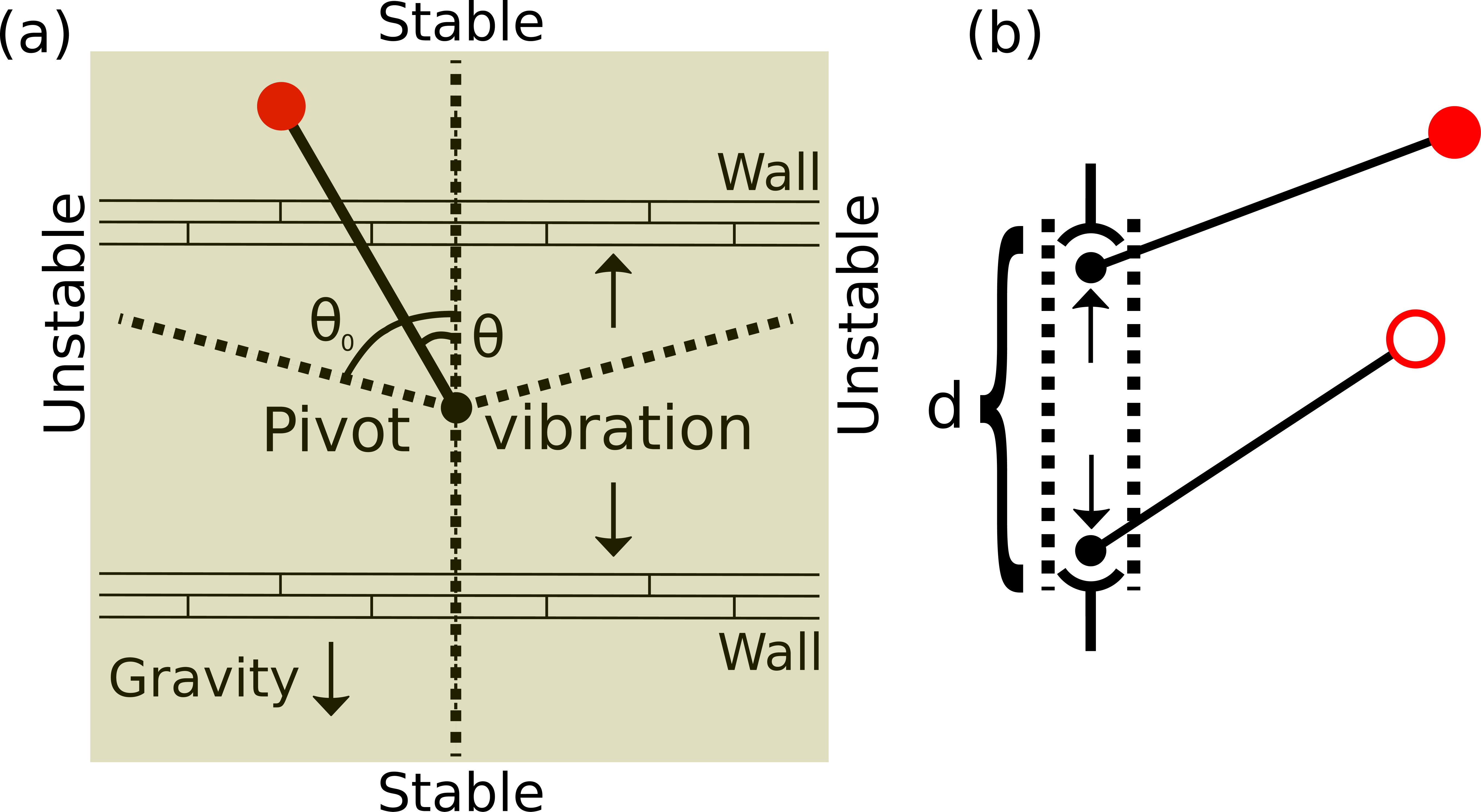}
    \caption{The same system as presented in Fig.\ \ref{fig:fig2}, but now additionally in a downward gravitational field. The unstable equilibrium positions now make an acute angle $\theta_0$ with the upward direction of the axis of pivot-vibration. When gravity is sufficiently strong the two unstable equilibrium directions merge with the axis, i.e. the top-equilibrium point loses stability.}
    \label{fig:fig5}
\end{figure*}
When gravity is gradually turned on, the position of the unstable equilibrium points change gradually.
The decrement $(\Delta\omega_{2n-1}-\Delta\omega_{2n})$ in angular velocity $\omega$ goes as ${\rm cos}(\theta)$ which is the derivative of the arm of the torque which in turn goes as ${\rm sin}(\theta)$. Hence the decrement $(\Delta\omega_{2n-1}-\Delta\omega_{2n})$ is more if $\theta$ is smaller. In presence of gravity, there should be an angle $\theta_0 < \pi/2$ where the decrement $(\Delta\omega_{2n-1}-\Delta\omega_{2n})$ during one full-cycle of pivot-vibration is undone by the anticlockwise increment due to the gravity during the same time (see Fig. \ref{fig:fig5}). For an angle $\theta < \theta_0$, the pivot vibration dominates and brings the oscillator back to vertically upward position. For $\theta < \theta_0$ gravity dominates. Hence the shifted unstable equilibrium positions (see Fig. \ref{fig:fig5}) are at $\pm\theta_0$ in presence of gravity, rather than $\pm\theta=\pi/2$. The stronger the gravity, the smaller is the angle $\theta_0$ where this condition gets satisfied. Above a certain strength of the gravity, $\theta_0$ reduces to $0$, and the top-equilibrium point loses its stability. 

One can of course look at the evolution of the equilibrium positions from the more familiar or chronological perspective where gravity is considered before the pivot vibration. From that perspective, the oscillator would initially have just the unstable top and stable bottom equilibrium points. When the pivot vibration is sufficiently fast, two unstable points of equilibrium emerge from the top-unstable equilibrium, and the top-unstable equilibrium itself turns stable. 

\section{Conclusion}
In this article I have provided a visually elaborate real-time pedagogical perspective to an age-old phenomenon, which has widespread and significant application in various fields of physics and nonlinear dynamics. I have also proposed a construction to create the phenomenon with a minimal drive, either in a passive form if dissipative forces can be eliminated, or in an active form, if dissipative forces can't be completely switched off. I used that minimal form of the drive to put forward the aforementioned visual perspective to the dynamics of the system. Additionally, the discussion presented here illustrates the number, nature, and evolution of the equilibrium positions of the system in absence and presence of gravity.  
\section*{Acknowledgement}
The author thanks Prof. Peter J. Hirschfeld and Prof. Andreas Kreisel for useful discussion and comments on the manuscript.
\section*{References}
\bibliography{aipsamp}
\end{document}